\pdfoutput=1

\pdfoutput=1

\documentclass[11pt]{scrartcl}
\usepackage[a4paper, total={16cm, 24cm}]{geometry}
\usepackage{microtype} 
\usepackage{authblk}
\title{Computational Social Choice:\\ Research \& Development}
\date{\vspace{-1.5cm}}

\author[1]{Dorothea Baumeister}
\author[2]{Ratip Emin Berker}
\author[3]{Niclas Boehmer}
\author[4]{Sylvain Bouveret}
\author[5]{Andreas Darmann}
\author[6]{Piotr Faliszewski}
\author[7]{Martin Lackner}
\author[8]{Jérôme Lang}
\author[9]{Nicholas Mattei}
\author[10]{Arianna Novaro}

\affil[1]{HS Bund, Federal University of Applied Administrative Science, Brühl, Germany}
\affil[2]{Carnegie Mellon University, Pittsburgh, United States of America}
\affil[3]{Hasso Plattner Institute, University of Potsdam, Germany}
\affil[4]{Univ. Grenoble Alpes, CNRS, Inria, Grenoble INP, LIG, Grenoble, France}
\affil[5]{University of Graz, Graz, Austria}
\affil[6]{AGH University of Krakow, Kraków, Poland}
\affil[7]{University of Applied Sciences St. Pölten, St. Pölten, Austria}
\affil[8]{CNRS, LAMSADE, Université Paris-Dauphine, PSL, Paris, France}
\affil[9]{Tulane University, New Orleans, United States of America}
\affil[10]{Université Paris 1 Panthéon-Sorbonne, CNRS, CES, Paris, France \vspace{0.25cm}}

\affil[ ]{\texttt{dorothea.baumeister@hsbund.de, rberker@cs.cmu.edu, niclas.boehmer@hpi.de, sylvain.bouveret@imag.fr, andreas.darmann@uni-graz.at, faliszew@agh.edu.pl, martin.lackner@ustp.at, jerome.lang@dauphine.fr, nsmattei@tulane.edu, arianna.novaro@univ-paris1.fr}}

\usepackage{xcolor}
\usepackage{graphicx}
\usepackage{pgfplots}
\pgfplotsset{compat=1.17}

\usepackage{amsmath, amssymb, amsthm, mathtools, nicefrac, siunitx}

\usepackage{thmtools}
\usepackage{thm-restate}

\theoremstyle{definition}

\usepackage[pagebackref]{hyperref}
\hypersetup{
	pdfencoding=auto,
	psdextra,
	colorlinks=true,
	citecolor=green!50!black,
	linkcolor=red!60!black,
}

\usepackage[nameinlink]{cleveref}

\usepackage{natbib}
\usepackage{algorithm}
\usepackage[noend]{algorithmic}

\usepackage[inline]{enumitem}
\usepackage{booktabs}
\usepackage{subcaption}

\usepackage[textsize=tiny]{todonotes}

\newcommand{\restatehere}[1]{%
	\marginline{\vspace{0.6cm}\footnotesize \hyperlink{original#1}{\hypertarget{restated#1}{[Main]}}}%
	\csname #1\endcsname*%
}

\begin{document}

\maketitle

\begin{abstract}
	\begin{center}
		\textbf{\textsf{Abstract}} \smallskip 
	\end{center}
	Computational social choice (COMSOC) studies principled ways to aggregate conflicting individual preferences into collective decisions.
    In this paper, we call for an increased effort towards \emph{Computational Social Choice: Research \& Development (COMSOC-R\&D)}, a problem-driven research agenda that explicitly aims to design, implement, and test collective decision-making systems in the real world.
    We articulate the defining features of COMSOC-R\&D, argue for its value, and discuss various roadblocks and possible solutions.
	
	\end{abstract}

\section{A COMSOC-R\&D Agenda}
The field of computational social choice (COMSOC) studies collective decision-making problems, where conflicting individual preferences must be aggregated into a collective outcome \citep{brandt2016handbook}. COMSOC offers a rich toolbox for achieving principled compromises, with applications ranging from participatory budgeting and fair resource allocation to group recommendations and deliberation.\footnote{Our use of the term COMSOC does not include work on matching under preferences 
(which studies school choice, course allocation, and kidney exchange problems, among others). While these topics are considered part of COMSOC by some, applied efforts in these domains have their origins largely in independent communities and have historically been much more driven by concrete problems and implementations (see, e.g., the survey by \citet{SonmezUnver2025}). We therefore consider such works rather role models for a broader vision of COMSOC-R\&D.}

Over the past 25 years, the COMSOC community has made remarkable progress in understanding the algorithmic and axiomatic foundations of collective decision-making problems. However, partly due to its roots in mathematics, theoretical computer science, and economics, the field has remained predominantly focused on foundational, abstract-level questions, whereas studies that develop, deploy, and evaluate mechanisms in specific real-world application contexts remain the exception rather than the norm.

This strong theoretical basis positions COMSOC to confront pressing real-world challenges in collective choice. However, to realize this potential, we need to complement COMSOC's foundational work with more application-driven research that bridges theory and practice. Accordingly, we call for a shift toward what we term \emph{Computational Social Choice: Research \& Development (COMSOC-R\&D)}, a research agenda that views social choice not only as an analytical framework but also as an applied engineering discipline for designing, implementing, testing, and deploying collective decision-making systems in the real world. 

\subsection{What is COMSOC-R\&D?}

COMSOC-R\&D aims to both apply existing social choice techniques and develop new ones in order to provide solutions for concrete, real-world problems. Rather than focusing primarily on developing new theoretical results, COMSOC-R\&D emphasizes the engineering and translational research goals of applying this research to the real world to yield implementable solutions.\footnote{We take some inspiration from the notion of \emph{translational research} in the biomedical fields, which aims to quickly move research out of the lab and into the real world, with the help of a translational partner, e.g., governments or businesses
\citep{wethington2015translational,collins2011reengineering}.} Accordingly, while a (standard) theoretical analysis might be part of it, a COMSOC-R\&D project places its main emphasis on translating theory into practice by implementing proposed algorithms or protocols with an accessible interface, evaluating them empirically, and ideally deploying them in real-world contexts. This process should also involve the release of software artifacts and code, following best practices from software engineering and other disciplines \citep{wilson2014best,wilson2017good}.

We would like to clarify that running computational experiments or developing software implementations of a general methodology is not sufficient to qualify for what we see as a proper COMSOC-R\&D approach. The distinctive feature of COMSOC-R\&D lies in its problem-driven orientation: every project should begin with a concrete problem in the real world and aim to progress through the entire cycle of design, implementation, testing, and application. This agenda draws inspiration from the distinction made between \emph{project-based} and \emph{product-based} engineering: while product-based initiatives are developed in anticipation of possible future demand and to be marketed later, project-based efforts are initiated in response to an existing, concrete client's need. Although the distinction between the two approaches is sometimes fuzzy, COMSOC-R\&D particularly seeks to encourage more project-based efforts  within COMSOC, which so far have remained comparatively rare.

\subsection{Why do We Need More COMSOC-R\&D?}\label{sec:why}

Any situation involving a group of agents who need to choose from a set of alternatives is a collective decision-making problem that could be addressed using COMSOC tools. In general, and unlike other areas of computer science, people facing such collective choice problems (like where to go for lunch or who to elect as a leader) can think of an \emph{ad hoc} solution, developed through experience or based on common sense. As a result, these decision-makers might not feel the need to turn to the scientific community, or they may not even be aware that their choices correspond to a social choice decision. This problem is shared with a broad set of communities that look at multi-criteria decision making (however, in particularly high-stakes settings, consultants or advisors are regularly employed \citep{WikiMCDA}). The resulting solutions for collective choice problems can often be sub-optimal, and developing or employing a better method requires scientific expertise that the COMSOC community can provide.

To illustrate this point, consider the example of participatory budgeting, where a municipality designates parts of its budget to be spent on projects proposed by citizens and selected according to a vote. In practice, the selection is often done by a simple greedy rule, which generally funds the projects receiving the most votes yet is provably suboptimal in terms of both fairness and efficiency.

Likewise, when the US National Science Foundation re-worked a mechanism to allocate telescope time, the resulting method was susceptible to strategic manipulation \citep{merrifield2009telescope}. 
Another example is the standard approach to \emph{Reinforcement Learning from Human Feedback}, which turns out to be equivalent to aggregating evaluators' preferences using the Borda count~\citep{siththaranjan2023distributional} and thus suffers from the same issues, such as manipulability by cloning~\citep{Conitzer24:Position,Berker25:From}.

In general, the deployment of COMSOC-R\&D work would bring numerous advantages to individuals and society. First, it is an opportunity for actual societal impact by improving how collective decisions are taken; in fact, in many applications, it could introduce collective decision-making in settings that did not involve active social participation before (as already achieved, e.g., in problems related to matching under preferences \citep{SonmezUnver2025}). 
Moreover, each successful application can bring more attention from the public to the full spectrum of problems that can be solved with COMSOC techniques. Given the current shortage of applied COMSOC work, additional COMSOC-R\&D efforts can offer valuable inspiration and guidance to researchers and practitioners facing similar problems, helping to build a shared knowledge base and best practices around how to conduct such research that can be adapted and reused. Consequently, successful COMSOC-R\&D projects can bring substantial recognition within the community.

Finally, and especially important from a research perspective, working on specific real-world problems typically surfaces aspects of the problem that are not adequately captured by models studied in theory. COMSOC-R\&D thus establishes a positive feedback loop between theory and practice: engineering abstract models to fit concrete settings reveals new constraints, objectives, and behavioral parameters, which, in turn, can inspire new models, axioms, and algorithmic questions in theoretical research \citep{mattei2021closing}---and also prevent misconceptions about the actual real-world relevance of some theoretical questions, or their importance in deployed systems.

\subsection{Successful Examples of COMSOC-R\&D}\label{sec:ex}

While COMSOC-R\&D currently constitutes only a very small fraction of COMSOC research, there are already several successful examples illustrating the kind of work we envision. In the following, we highlight three success stories that showcase different aspects of COMSOC-R\&D. There are, of course, many more, 
including work on peer reviewing \citep{merrifield2009telescope,meir2021market,rozencweig2023mitigating}, designing conference timetables \citep{ReyEndrissSchedule2024}, allocating classrooms \citep{kurokawa2018leximin}, managing resources for Earth observing satellite missions \citep{lemaitre2003equitable,dyer1976actual}, and allocating food bundles \citep{benade2024achieving}.

\paragraph{Participatory Budgeting}
The Method of Equal Shares (MES)~\citep{peters20, NEURIPS-Peters} was developed as a theoretical proof of concept that in participatory budgeting, groups of voters can be represented proportionally to their size. A few cities recently adopted it, addressing concrete issues they were facing with the widely used greedy method, such as a well-coordinated group of citizens hijacking a large fraction of the budget. MES quickly turned into a COMSOC-R\&D effort.\footnote{See \href{https://equalshares.net/}{https://equalshares.net/}.} It is highly configurable, and different design choices can strongly affect its nature, such as selecting more or fewer low-cost projects~\citep{FaliszewskiFPP023}. Tools have been developed to allow decision makers to experiment with the rule\footnote{\url{https://pref.tools/pabutools/}} and to present the results intuitively \citep{boehmer2024evaluation}.

\paragraph{Sortition} Another recent success story and example of COMSOC-R\&D is the development of methods and tools for composing citizens' assemblies that are both representative and (individually) fair. The Panelot platform\footnote{\href{https://panelot.org/}{https://panelot.org/}} has been developed to implement sortition-based methods that fulfill these two requirements. The underlying work, driven by concrete needs, has already been used on several occasions to create actual assemblies \citep{FlaniganG0HP21}.

\paragraph{Nominated Proof-of-Stake Blockchains} A deployed application of multiwinner voting \citep{FSST-trends,abcbook}---and an example of successful COMSOC-R\&D---is the nominated proof-of-stake system used by Polkadot \citep{burdges2020overview,blockchain2/AbbasCP22}. For this application, a tailored voting mechanism was developed, drawing on previous foundational work from the COMSOC community. The desired voting method had to comply with desiderata that had previously received little or no attention (e.g., over-representation, scalability, verifiability) and thus led to important general algorithmic developments \citep{CevallosS21,mathprog-phragmen,FernandezGFB24,boehmer2024approval}.

\section{Roadblocks and Possible Solutions}\label{sec:how}

Although there seems to be broad agreement within the community that COMSOC-R\&D work is highly beneficial (see \Cref{sec:why}), such work remains comparatively rare (see \Cref{sec:ex}). We discuss possible reasons, distinguishing general challenges common to applied computer science research from what we see as the most pressing issues more specific to COMSOC-R\&D (though not exclusively).
The goal of this section is to raise awareness of potential roadblocks to COMSOC-R\&D projects and to describe solution strategies, both at the individual and the community level.

\subsection{Common Roadblocks}

There is a rich literature on obstacles that academic computer science researchers face when trying to translate and apply their research \citep{DBLP:journals/corr/abs-2506-14829,DBLP:conf/aies/BondiXAK21,DBLP:journals/see/FloridiCKT20, Daun-SE,Amato2025,Agapie2024}. In addition, both the Computing Research Association\footnote{\url{https://cra.org/resources/best-practice-memos/}} and Nature publishing\footnote{\url{https://www.nature.com/ncomms/submit/applied-science-research}} have issued best practice guidelines for starting, conducting, and reviewing applied science and engineering work, including work for social good and social change \citep{hager2019artificialintelligencesocialgood}.
We highlight three general roadblocks that we believe are particularly salient for COMSOC.

\paragraph{Partnering} Building and sustaining links between researchers and real-world stakeholders requires an enabling network and substantial time. Misaligned timelines and values, limited mutual awareness, and the upfront work of establishing shared terminology and realistic expectations further complicate collaboration. At the community level, we can facilitate partnerships by creating venues and discussion forums where both academics and practitioners are regularly present. While still rare, there are emerging efforts to build such systemic bridges, for example, the EDDY network\footnote{\href{https://www.eddy-network.eu/}{https://www.eddy-network.eu/}}. Matchmaking platforms and forums like industry days have proven useful in other CS communities \citep{holstein2019improving,garousi2016challenges}, but they require continuous maintenance efforts and sustained awareness on both sides. At the individual level, a common success strategy is to jointly search for grant opportunities, which creates buy-in from all parties.

\paragraph{Incentives} Researchers may doubt the ease of publishing academic articles of R\&D-style work when theoretical novelty takes a back seat to engineering and deployment. Indeed, conventional evaluation rubrics often undervalue adaptation efforts, carefully reasoned design trade-offs, and negative or mixed results. Additionally, more engineering-focused work tends to attract fewer citations, as it often prioritizes context-specific design over generalizable methodology. 

The incentive problem calls for community-wide action. Established venues need to create space for such papers, for instance, through dedicated tracks and evaluation criteria. Some conferences, such as AAAI and IJCAI, already have both social impact tracks and deployed applications tracks; while there is not yet a strong tradition of COMSOC-R\&D being published there, these tracks are well-aligned with the scope we are advocating for. At venues without dedicated tracks, a central challenge is that reviewers may find it difficult to evaluate the quality of COMSOC-R\&D submissions, given that these are still ``non-standard'' papers within COMSOC, and standard evaluation criteria are only partly applicable. As a first step towards addressing this, we propose in \Cref{sec:rev} a set of review criteria tailored to COMSOC-R\&D research.

\paragraph{Operations} There is a huge gap between developing a proof-of-concept research software and a full-fledged tool intended for use by a large number of people \citep{wilson2014best,wilson2017good}. Building a full-fledged system requires developing an intuitive interface, deploying a scalable application on the web, managing security issues, and maintaining this infrastructure over a sustained period of time. Without special care for this invisible and undervalued work (within the academic ecosystem), a development project may fail. One particular hurdle related to maintenance is ensuring that the stakeholders are able to continue their operations even after the research team moves on to other
projects. A high-effort way of achieving this is by starting a spin-off company, but doing so is not always reasonable (see the Decision Analysis area of INFORMS for successful examples \citep{olson1997decision}\footnote{\url{https://pubsonline.informs.org/magazine/orms-today/2020-decision-analysis-software-survey}}). Fortunately, there are often other solutions. For example, in the case of participatory budgeting and MES, one city hired an external software company to integrate MES into its system, thereby ensuring independent support.

\subsection{COMSOC-Specific Roadblocks}
We now turn to more COMSOC-specific challenges; while some of these are shared by other areas, we find these challenges to be more salient for the COMSOC research community. To avoid discouragement, we also highlight potential remedies at both the individual and community level, alongside each of these challenges.

 \paragraph{High-stakes Decisions} Many COMSOC applications are particularly high-stakes and closely scrutinized (e.g., democratic elections), which can erode researchers’ confidence. The lack of ``perfect'' solutions and the resulting need to balance competing normative criteria further increase the burden of justification and the risk of stakeholder push-back, a problem shared with many multi-criteria decisions \citep{WikiMCDA}. A general mitigation strategy to adequately address the high-stakes nature inherent to collective decision-making problems is to start with small problems in a local environment, building upon personal connections. 
 For instance, the success of the Method of Equal Shares started with Wieliczka in Poland, a small city on the outskirts of Krakow. The mutual trust between the MES team and the city was based on previous personal contacts. Yet, even in this case, the city adopted MES only in an experimental ecology-oriented participatory budgeting exercise. However, after this initial success, other cities became more open to adopting MES. Even if no such obvious small-scale variant exists, at the level of broadly applied high-impact problems, there begins to be room to develop trial-run solutions, provided that expectations are managed accordingly. Further, at the community level, we can facilitate more “work-in-progress’’ forums for projects and ideas targeting real-world challenges, to provide low-threshold exchange mechanisms and to share responsibility.
 
\paragraph{Lack of Clear Best Practices} Visible role models, documented success stories, and codified best practices for COMSOC-R\&D remain scarce. With every additional COMSOC-R\&D paper, this problem will diminish. Still, there are more concrete efforts the community can undertake, such as highlighting success stories (as in \Cref{sec:ex}), conducting meta-studies, publishing best practices, and, when writing about specific COMSOC-R\&D projects, explicitly describing not only the results but also the process and lessons learned, so that others in the community can build on them (cf.\ \Cref{sec:rev}). Taking these steps has led to success in other application-oriented parts of computer science \citep{Agapie2024}, including software engineering \citep{garousi2016challenges}.

\paragraph{Culture Gap} COMSOC researchers, accustomed to rigorous mathematical arguments and formalisms, are often less comfortable with and less passionate about the rather complicated specifications, ambiguous requirements, and trade-off decisions inherent to real-world projects. In particular, theoretical papers often address one specific issue at a time, whereas engineering work mixes several of them.
Training settings would be highly valuable to address culture gaps and the associated lack of experience. One idea is a recurring challenge series (analogous to the PACE challenge in parameterized complexity\footnote{\href{https://pacechallenge.org/}{https://pacechallenge.org/}}), in which a real-world problem is posed to the community, and submitted mechanisms and ideas are evaluated by real-world stakeholders. This could, in turn, contribute to increasing the visibility of COMSOC-R\&D. 
    
\paragraph{Status Quo Bias} Stakeholders may often be attached to existing, sometimes autocratic, procedures: as long as these are perceived as ``working'' and those in control are content with the outcomes, there is little incentive to replace them with more principled mechanisms. A natural response is outreach that clearly communicates the pitfalls and complexity of collective decision-making. Raising public awareness of specific mis-handlings of decision processes can also increase pressure for change, but typically comes with confrontation and must therefore be approached with care. The community can redouble efforts for publicity and outreach along the main thrusts of collective decision-making.

\paragraph{Lack of Trust} Stakeholders outside academia may view researchers as idealistic theoreticians who do not understand practical constraints and difficulties and who lack real-world experience. One approach to address their concerns is to show that the researchers can ``speak the same language,'' i.e., they are aware of various practical issues and possible problems. Building this competence requires effort and experience; this is another reason why starting with small-scale problems in the local environment is important. Here we can borrow from work in the human-computer interaction, design, and AI for social good communities, focusing on participatory design and methodologies of building long-term trust and engagement within communities \citep{abebe2020roles,birhane2022power,hager2019artificialintelligencesocialgood}.

\section{Merits of COMSOC-R\&D Papers and Proposed Reviewing Guidelines} \label{sec:rev}

As we have argued, COMSOC-R\&D work faces several obstacles. One specific (but academically important) issue is the difficulty of publishing engineering-heavy work in a theory- and methods-inclined community. As a first step toward addressing this, we argue that COMSOC-R\&D work provides \textit{substantial benefits to the development of the field as a whole} (cf.\ \Cref{sec:why}).
To highlight this value, we now outline aspects that are especially important when writing COMSOC-R\&D papers, which should also be taken into account when evaluating and reviewing them. We do not expect every COMSOC-R\&D paper to address every single one of these points; rather, when writing or reviewing these applied papers, one should weigh these factors in addition to (or in place of) more traditional criteria for evaluating theoretical contributions.

    \textbf{Modeling \& Motivation}
    Any COMSOC-R\&D work requires substantial effort in modeling a real-world scenario. The description of the model and the underlying rationale are a major part of the work. Since the application determines the model (and not the other way around), it cannot be expected that these models are as `elegant' as those typically studied in theory. Evaluation should therefore focus less on the model and its appeal in isolation and more on the application itself and the fidelity of the model \emph{to that application}, including key features and constraints.
     
    \textbf{Literature} 
    An important feature of COMSOC-R\&D work is its connection to the COMSOC literature. In practice, the proposed solution might only be loosely related to previous work. Still, discussing related work -- where it was useful, where it failed -- is highly valuable: If the proposed solution is firmly grounded in previous (more theoretical) publications, it is a success for our community, and if not, this gap becomes a call for further research.
    
    \textbf{Empirical Assessment} 
    A central contribution of COMSOC-R\&D work lies in evaluating proposed mechanisms in realistic settings. This includes selecting and explaining meaningful evaluation metrics, which may differ substantially from established theoretical desiderata or even contradict them. Comparisons with baseline methods (including the status quo) are valuable, in particular if they highlight the effect of approaching the problem from the perspective of collective decision-making.
    
    \textbf{Data} 
    Considerable effort has been put into collecting data sets suitable for COMSOC research \citep{preflib,FaliszewskiFPP023,BoehmerS23}. Additional data sets from new applications are extremely valuable, especially as recent work \citep{SzufaBBFNSST25-map} has shown that preference data can differ substantially depending on the domain. 
     
    \textbf{Transparency}  
    Engineering work is inherently iterative. Papers should therefore document the key design steps, discussions, and lessons learned during the project as valuable guidance for future COMSOC-R\&D projects. Indeed, many small modeling choices and fine-tuning decisions must be made. While these will often need to be \emph{ad hoc} and cannot all be justified in depth, it is crucial to be transparent and explicit about them.
     
    \textbf{Reproducibility} 
    The full scope of COMSOC-R\&D projects may not be reproducible. This can be due to physical impossibilities (the engineering project under consideration cannot be repeated under the same conditions) or contractual obligations (it is necessary to withhold key data sets, software). Nevertheless, authors should aim for reproducibility wherever possible. Providing an easily accessible interface or artifacts that others can build on is a strong plus.
    
    \textbf{Stakeholder Engagement} 
    COMSOC-R\&D projects typically involve sustained collaboration with non-academic partners. Documenting this process -- how requirements were elicited, how design decisions were made, and how feedback was incorporated -- provides valuable guidance for future projects.
    
    \textbf{Implementation and Deployment} 
    The practical challenges of deploying a collective decision-making system (user interface design, integration with existing processes, computational optimization, etc.) are often substantial and rarely addressed in theoretical work. Detailed accounts of such deployment efforts are a distinctive contribution of COMSOC-R\&D papers.

Incorporating these perspectives into the reviewing process does not necessarily require a change in reviewer guidelines. Rather, this is a call for consideration to the (wider) COMSOC community:  COMSOC-R\&D papers are valuable for the community, and hence should be judged by their merit rather than by how closely they resemble typical COMSOC papers. To give COMSOC-R\&D papers an equal standing in the reviewing process, it has to be acknowledged that their merit differs from that of theory-focused papers.

\section{A Vision for the Future}

We envision a future where COMSOC-R\&D papers will be equally represented at COMSOC venues alongside more classical, theory-focused papers. We believe that the COMSOC-R\&D approach has great potential to bring concrete improvements in the way real-life decisions are made every day.  Particularly promising application domains include \emph{preferences-aware timetabling} (e.g., for university classes or work shifts; cf. \citep{dighum-schedules}), \emph{dynamic reallocation of objects under uncertainty} (e.g., for reassigning offices during a building renovation), and \emph{fair public transport}.
We invite researchers to pursue these challenges and call on the community to recognize and reward such efforts.

\subsection*{Acknowledgments}

This work was initiated in a work group during the
\href{https://www.dagstuhl.de/de/seminars/seminar-calendar/seminar-details/25401}
{Dagstuhl Seminar 25401} ``Societal Impact of Computational Social Choice''.
This project has received funding from the European Research Council (ERC) under the European Union’s Horizon 2020 research and innovation programme (grant agreement No 101002854).

\begin{center}
\includegraphics[width=3cm]{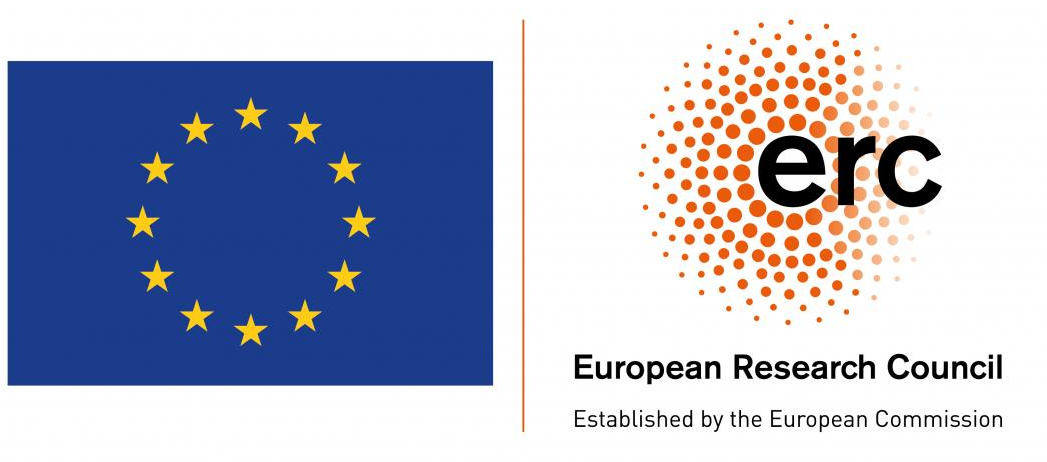}
\end{center}


\end{document}